\documentclass{aa}
\usepackage{graphics}
\usepackage{times}

\newcommand{\unoe}  {1E 1740.7$-$2942}

\newcommand{\grp}   {${\rlap.}^{\circ}$}
\newcommand{\pri}   {${\rlap.}^{\prime \prime}$}
\newcommand{\rl}    {${\rlap.}^{s}$}

\newcommand{\ltsima} {$\; \buildrel < \over \sim \;$}
\newcommand{\simlt}  {\lower.5ex\hbox{\ltsima}}            
\newcommand{\gtsima} {$\; \buildrel > \over \sim \;$}
\newcommand{\simgt}  {\lower.5ex\hbox{\gtsima}}            

\begin{document}

\thesaurus{08(08.09.2 \object{\unoe}; 13.09.6; 13.18.5; 13.25.5)}

\title{VLT search for the infrared counterpart of \unoe}

\author{Josep Mart\'{\i}\inst{1} 
\and    I. F\'elix Mirabel\inst{2,3}
\and    Sylvain Chaty\inst{4}
\and    Luis F. Rodr\'{\i}guez\inst{5}
}

\institute{
Departamento de F\'{\i}sica, Escuela Polit\'ecnica Superior,
Universidad de Ja\'en, Calle Virgen de la Cabeza, 2, E-23071 Ja\'en, Spain
\and
CEA/DSM/DAPNIA/Service d'Astrophysique, Centre d'\'Etudes de Saclay,
F-91191 Gif-Sur-Yvette, France
\and
Instituto de Astronom\'{\i}a y F\'{\i}sica del Espacio, C.C. 67, Suc. 28, 1428 Buenos Aires, Argentina
\and
Physics Department, The Open University, Walton Hall, Milton Keynes, MK7 6AA, United Kingdom
\and
Instituto de Astronom\'\i a, UNAM, Campus Morelia, Apdo. Postal 3-72, 58089 Morelia, Michoac\'an, M\'exico
}

\offprints{J. Mart\'{\i}, jmarti@ujaen.es}

\date{Received / Accepted}

\maketitle

\begin{abstract}

We report the results of our search for the near infrared
counterpart of the microquasar \unoe\ using the VLT\footnote{Based on observations collected
at the European Southern Observatory, Chile (ESO No 63.H-0261).}.
For the first time, several counterpart candidates have been found in our $Ks$-band images that may
be consistent with the best radio and X-ray positions available for \unoe.
However, the non-detection of variability between two observing epochs and the
positional uncertainty still remaining at the
\ltsima$1^{\prime\prime}$ level prevent us from identifying an unambiguous
counterpart. Alternatively, the VLT images set new upper limits significantly deeper than previously reported
that constrain the binary companion to be later than B8 V or earlier than G5 III.


 
\end{abstract}

\section{Introduction} \label{intro}

The Galactic Center (GC) in hard X-rays is dominated by the two bright sources
\unoe\ and GRS 1758$-$258 (Sunyaev et al. 1991;   Goldwurm et al. 1994). In spite of nearly one decade
of observation, the true nature of these objects still remains a mystery.
Nevertheless, from their likely radio counterparts with bipolar jets 
(Mirabel et al. 1992; Rodr\'{\i}guez et al. 1992)
they are widely accepted to be microquasar systems with persistent activity
at both X-ray and radio wavelengths. The reader is referred to the recent
review by Mirabel \& Rodr\'{\i}guez (1999) for an updated account.

Accepting the radio counterpart identification, the position of the two GC microquasars  
is known to better than one arcsecond thanks to the Very Large Array (VLA)
interferometer in its most extended configurations. Even with this
knowledge in hand the search for an optical/infrared counterpart in the \unoe\ case 
has proven to be virtually impossible. 
The best upper limits previously available were 
$z>22$-23 (Mereghetti et al. 1992),  
$I>21$ mag (Leahy et al. 1992) and 
$K>17-18$ (Djorgovski et al. 1992; Chaty et al. 1998) 
Two circumstances conspire against a successful detection. 
First, the huge interstellar absorption ($A_V \geq 50$ mag) 
in the line of sight to \unoe.
Second, the richness of the field especially in the $J$ and $K$-bands. 
These difficulties also yield most optical astrometric standards practically
unrecognizable in infrared frames. The situation seems to be a little bit less severe
for GRS 1758$-$258, where a few possible candidates close to the VLA position have been
proposed (Mart\'{\i} et al. 1998).

In this note we present the $Ks$-band images of the \unoe\ field taken with the 
first 8.2 m unit (Antu) of the Very Large Telescope (VLT) of the European Southern Observatory (ESO). 
The VLT images are significantly deeper than the upper limits reported
in previous infrared searches. For the first time, we are able
to detect some counterpart candidates that may be consistent with the accurate radio and X-ray positions. 
Although an unambiguous identification is not yet possible, the first VLT images of the
\unoe\ field represent a useful template against which future observations can be compared.

\section{Observations}

We observed \unoe, in service mode, on two different epochs using the ISAAC instrument of 
the VLT in imaging mode (see Cuby 1999 for a description). In both occasions,
the $Ks$ filter centered at the 2.16 $\mu$m wavelength was used. 
The first epoch was on 1999 April 04 (JD 2451272.834), with a total exposure time of 3000 s under windy conditions. 
The second observation took place on 1999 May 02 (JD 2451300.873), with a total exposure time
of 1400 s under a more quiet atmosphere. Both nights
were considered of photometric quality, but the seeing in May 02 (0\pri 5) was much better than in April 04 (0\pri 7).
The Julian dates given correspond to the middle of the observation.
The data were reduced using the IRAF package, including sky background subtraction, 
flat field division and frame combination, and also photometry in crowded fields was performed thanks to the
DAOPHOT package.
The photometric zero point was determined using the magnitudes of several
stars in the field that were known from many previous observing runs with 
different ESO telescopes in La Silla (Chaty 1998). 
  

VLT-ISAAC observations were also conducted at longer wavelengths using
the $L$-band (3.78 $\mu$m) filter on two nights in 1999 June. Unfortunately, both $L$-band epochs suffered
from either non-photometric conditions or too bad seeing and they appear to be of little use.
Therefore, we will concentrate our analysis on the $Ks$ images only.

\section{Results and discussion}

\begin{figure}[htb]
\mbox{}
\vspace{14.0cm}
\includegraphics{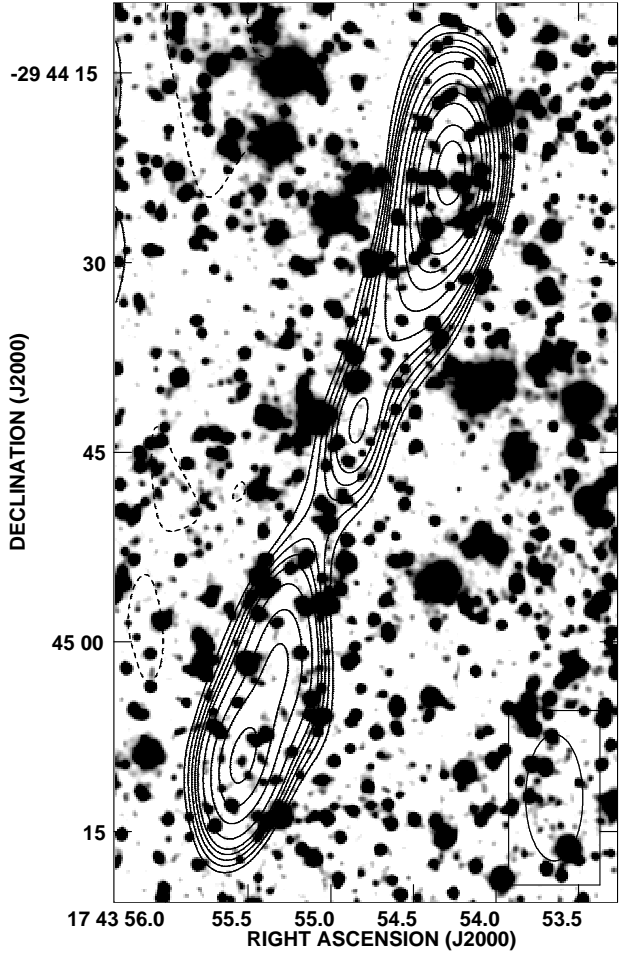}
\caption[]{VLT+ISAAC image of the \unoe\ field using the $Ks$ filter.
The solid contours represent the microquasar radio emission at 6 cm and they correspond
to $-3$, 3, 4, 5, 6, 7, 9, 12, 15, 18, 22, 30 and 35 times 20 $\mu$Jy beam$^{-1}$, the rms noise.
The synthesized beam is shown as an ellipse at the bottom right corner measuring
10\pri 0 $\times$ 4\pri 5, with position angle 0\grp 0.
}
\label{isaac}
\end{figure}

A wide view of the \unoe\ field in the $Ks$ infrared band
is presented in Fig. \ref{isaac}. This figure contains the $Ks$ image
of the second VLT epoch with the VLA 6 cm data overlayed on it for illustrative purposes. 
The radio contours display the appearance
of the microquasar bipolar jets emanating from a central core against the near infrared background.
No special increase, nor decrease, in the infrared star counts seems to be present in association with
the arcminute extended radio lobes.
 
\begin{figure*}[htb]
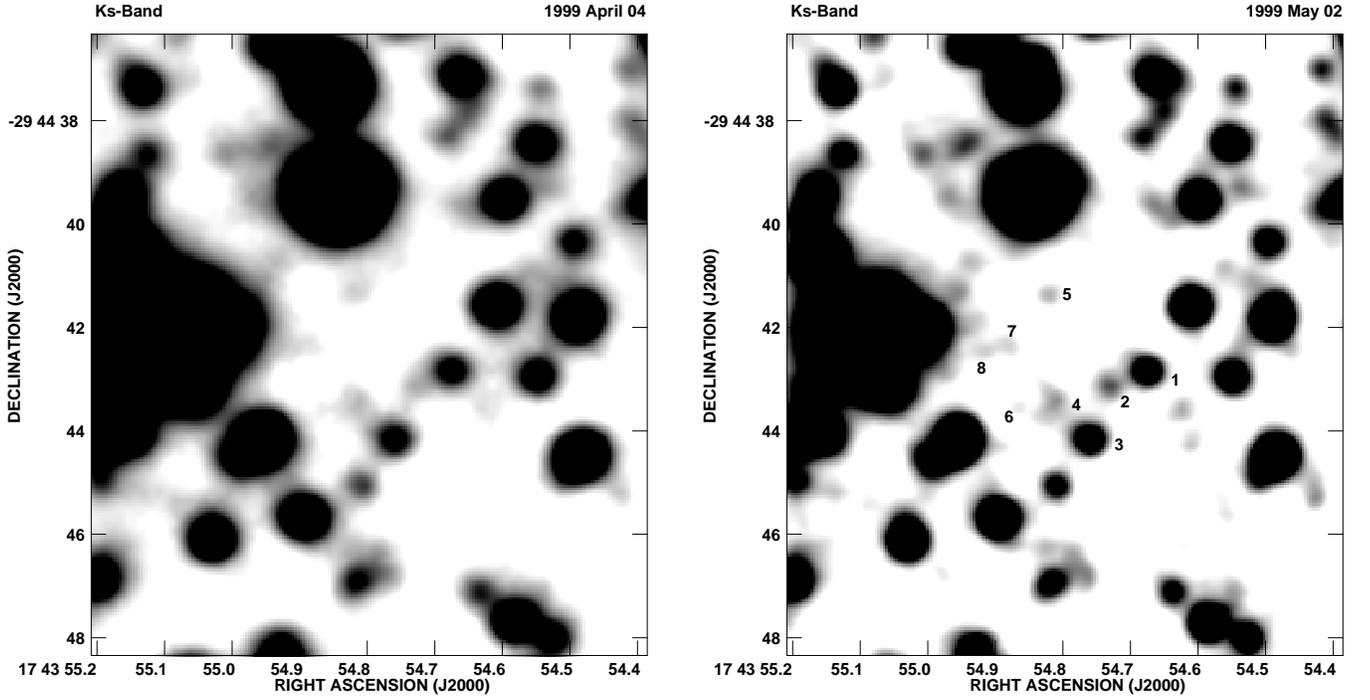

\mbox{}
\vspace{9.0cm}
\includegraphics{jmarti.f2a}
\includegraphics{jmarti.f2b}
\caption[]{A close up view of the \unoe\ field with the VLT+ISAAC on two different epochs
at the 2.16 $\mu$m wavelength. 
The infrared sources newly detected in its vicinity are labelled from 1 to 8.
The corresponding $3\sigma$ upper limit of both panels is $K=19.5$ mag. 
}
\label{2epoch}
\end{figure*}

A close up view of the central arcseconds around the \unoe\ central core 
is presented in Figs. \ref{2epoch}a and \ref{2epoch}b
for both the first and second epoch, respectively. The 
astrometric solution on these Figures has been obtained
by using 7 unsaturated USNO-A V2.0 stars, assumed to be 
reliably identified on the VLT frames.
The corresponding residuals of the astrometric fit are about $\pm$0\pri 20, thus suggesting
that the identification of the reference stars is indeed correct. Therefore,  
our astrometry should be reliable within a $3\sigma$ error of $\sim0$\pri 60 when
plotting X-ray or radio positions on the VLT grid. Such a conservative approach is 
justified given the difficulties of performing good astrometry in the crowded and heavily absorbed field 
of \unoe. 
Several new infrared sources detected 
have been labelled from 1 to 8. 



\begin{figure}[htb]
\mbox{}
\vspace{8.0cm}
\includegraphics{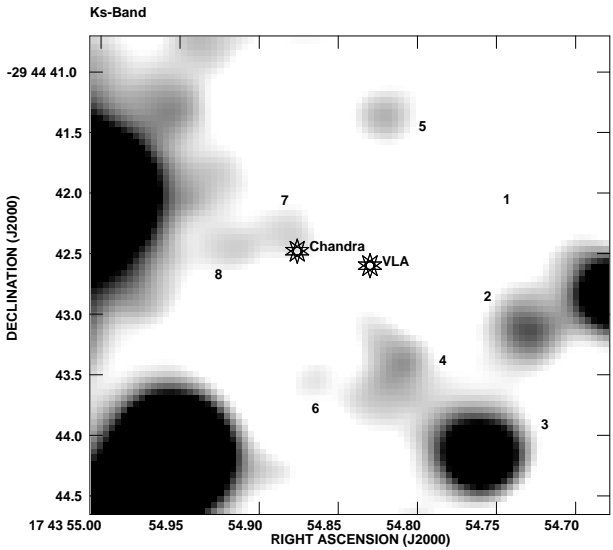}
\caption[]{Expanded view of the second epoch image. The VLA radio and Chandra X-ray
positions have been plotted as star symbols whose size is comparable to
their respective formal uncertainty. The residual 0\pri 6 offset between them is within
the Chandra absolute positional error, as expected for both the X-ray and radio source being the same object. 
}
\label{zoom}
\end{figure}

For detailed discussion purposes,
an expanded view of our second epoch image is shown in Fig. \ref{zoom}. Here, the most accurate
radio and X-ray positions available have been plotted according to our astrometry.
At radio wavelengths we are using the VLA position
$\alpha_{\rm J2000}=  17^h 43^m 54$\rl 83  and
$\delta_{\rm J2000}= -29^{\circ} 44^{\prime} 42$\pri 60
with $\pm$0\pri 1 accuracy in each coordinate. This position is slightly
different from that reported in Rodr\'{\i}guez et al. (1992) since it has been derived
from the expanded VLA data set, as shown in Fig. \ref{isaac}.
Haller \& Melia (1994) have published a $K$-band image of the \unoe\ field with astrometric information,
based on 8 reference stars,
taken with the 2.3m telescope at the Steward Observatory (limiting magnitude $K=16.5$).
It is reassuring that
the VLA position seems to be consistent in both cases with respect to nearby bright infrared sources.

In X-rays, the best source location currently available
is provided by the Chandra X-Ray Observatory being 
$\alpha_{\rm J2000}=  17^h 43^m 54$\rl 876$\pm0$\rl 004 and 
$\delta_{\rm J2000}= -29^{\circ} 44^{\prime} 42$\pri 48$\pm$0\pri 04, 
where the errors represent roughly 90\% confidence limits (Cui et al. 2000).
The Chandra position is offset 0\pri 6 from the VLA coordinates, which is within the uncertainty in
the Chandra absolute aspect solutions (Cui et al. 2000).


By inspecting the VLT images, we detect at least 8 sources in the vicinity of the X-ray and radio position.
We label them from 1 to 8 in Fig. \ref{2epoch}b, the one with best seeing.
Their $Ks$-band magnitudes are listed in Table \ref{mag}. The magnitude for object 8 is
very uncertain because of severe background contamination.
Among these new sources, those labelled as 4, 6, 7 and 8 are the ones whose coincidence with
the VLA and Chandra positions cannot at present be ruled out. This is because of the
\ltsima$1^{\prime\prime}$ uncertainties still remaining in
the absolute positioning of the Chandra and VLA positions on the VLT image.
Interestingly, source 4 looks slightly different than the seeing profile and may have extended appearance.
No photometric variability above the quoted magnitude errors was observed between the two
observing epochs. This fact, together with the presence of several candidates, still prevents us
from proposing a reliable counterpart candidate based on present data.

\begin{table}
\caption[]{\label{mag} Photometry of the candidate counterparts}
\begin{tabular}{ccc}
\hline
Object   & $Ks$      & Remarks   \\
         & (mag)    &           \\
\hline
1        & $16.7 \pm 0.1$  &                      \\
2        & $18.2 \pm 0.2$  &                      \\
3        & $16.6 \pm 0.1$  &                      \\
4        & $18.0 \pm 0.1$  & Possibly extended    \\
5        & $18.4 \pm 0.3$  &                      \\
6        & $19.2 \pm 0.3$  &                      \\
7        & $18.7 \pm 0.3$  &                      \\
8        & $19   \pm 1$    & Uncertain background \\
\hline
\end{tabular}
\end{table}

If one of the astrometrically selected sources is indeed the correct \unoe\ counterpart,
we can adopt $K=18.5$-$19.0$ for the infrared magnitude of the system. Given the
similarity in hard X-rays of \unoe\ and the classical black hole candidate Cygnus X-1,
it is instructive to derive which is the interstellar absorption required for both objects
having the same infrared luminosity. The absolute $K$-band magnitude of Cygnus X-1
is $K=-5.8\pm0.2$ (Beall et al. 1984). At the Galactic Center distance of 8.5 kpc,
a Cygnus X-1 system would appear 
with the magnitudes observed for the candidates, provided that an absorption
of $A_K=9.7$-$10.2$ magnitudes exists. According to usual extinction laws
(Rieke \& Lebofsky 1985; Predehl \& Schmitt 1995) this implies a visual absorption of
$A_V=86$-91 magnitudes, equivalent to a hydrogen column density of
$N_H = (1.5$-$1.6) \times 10^{23}$ cm$^{-2}$. This value is perfectly consistent with the 
interstellar absorption from X-ray spectral fits, that provide 
$N_H > 8 \times 10^{22}$ cm$^{-2}$ (Churazov et al. 1996; Sheth et al. 1996).

If none of sources 4, 6, 7 and 8 is the true \unoe\ counterpart, the 
$3\sigma$ limiting magnitude
of our best image ($Ks=19.5$) implies that the absolute magnitude is $K>-0.1$. In this case,
the stellar types allowed would be main sequence stars later than about B8 V or giants earlier than
G5 III. 

In any case, further observations will be necessary to finally 
detect the \unoe\ counterpart and establish its true nature.

\begin{acknowledgements}
JM acknowledges partial support by DGICYT (PB97-0903)
and by Junta de Andaluc\'{\i}a (Spain). IFM acknowledges financial support from CONICET/Argentina.
SC acknowledges financial support from grant F/00-180/A from the
Leverhulme Trust.
LFR acknowledges the support from DGAPA, UNAM and CONACyT, M\'exico.
We also thank S. Mereghetti and T. Belloni for useful comments.

\end{acknowledgements}


\begin{thebibliography}{}

\bibitem[]{} Beall J.H., Knight F.K., Smith H.A., et al., 1984, ApJ, 284, 745

\bibitem[]{} Chaty S., 1998, PhD Thesis, University of Paris Sud XI

\bibitem[]{} Churazov E. et al., 1996 ApJ 464, L71 

\bibitem[]{} Cuby J.G., 1999, {\it ISAAC User Manual}, ESO

\bibitem[]{} Cui W., Schulz, N.S., Baganoff F.K., Bautz M.W., Doty J.P., et al., 2000, ApJ (submitted)

\bibitem[]{} Djorgovski, S., Thompson D., Mazzarella J., Klemola A., 1992, IAU Circ. 5596

\bibitem[]{} Goldwurm A., Cordier B., Paul J., et al., 1994, Nat, 371, 58

\bibitem[]{} Haller J.W., Melia F., 1994, ApJ, 423, L109

\bibitem[]{} Leahy D.A., Langill P., Kwok S., 1992, A\&A, 259, 209

\bibitem[]{} Mart\'{\i} J., Mereghetti S., Chaty S., Mirabel I.F., Rodr\'{\i}guez L.F., 1998, A\&A, 338, L95

\bibitem[]{} Mereghetti S., Caraveo P., Bignami G.F., Belloni T., 1992, A\&A, 259, 205  

\bibitem[]{} Mirabel I.F., Rodr\'{\i}guez L.F., Cordier B., Paul J., Lebrun F., 1992, Nat, 358, 215

\bibitem[]{} Mirabel I.F., Rodr\'{\i}guez L.F., 1999, ARAA 37, 409 

\bibitem[]{} Predehl, P.,  Schmitt, J., 1995, A\&A, 293, 889

\bibitem[]{} Rieke, G., Lebofsky, M., 1985, ApJ, 288, 618

\bibitem[]{} Rodr\'{\i}guez L.F., Mirabel I.F., Mart\'{\i} J., 1992, ApJ, 401, L15


\bibitem[]{} Sheth S. et al. 1996 ApJ 468, 755 

\bibitem[]{} Sunyaev, R.A., et al., 1991, A\&A, 247, L29


\end{thebibliography}
\end{document}